\documentclass{sig-alternate}
\usepackage[outdir=./]{epstopdf}
\usepackage{url}
\newcommand{\subparagraph}{}
\begin{document}

	\title{SourcererCC and SourcererCC-I: Tools to Detect Clones in Batch mode and During Software Development}
	\author{\alignauthor Vaibhav Saini, Hitesh Sajnani, Jaewoo Kim, and Cristina Lopes \\
		\affaddr{Dept. of Information and Computer Science, University of California} \\ \affaddr{ Irvine, CA, USA} \\
		\email{vpsaini@uci.edu, hsajnani@uci.edu, jaewoo@uci.edu, lopes@uci.edu}
	} 
	\maketitle
	
	\begin{abstract}
		Given the availability of large source-code repositories,
		there has been a large number of applications for large-scale clone detection. Unfortunately, despite a decade of active
		research, there is a marked lack in clone detectors that scale
		to big software systems or large repositories, specifically for detecting near-miss (Type 3) clones
		where significant editing activities may take place in the
		cloned code.
		
		This paper demonstrates: (i) SourcererCC, a token-based clone detector
		that targets the first three clone types, and exploits an index
		to achieve scalability to large inter-project repositories
		using a standard workstation. It uses an optimized
		inverted-index to quickly query the potential clones
		of a given code block. Filtering heuristics based on token ordering
		are used to significantly reduce the size of the index,
		the number of code-block comparisons needed to detect the
		clones, as well as the number of required token-comparisons
		needed to judge a potential clone; and (ii) SourcererCC-I, an Eclipse plug-in, that uses SourcererCC's core engine
		to identify and navigate clones (both inter and intra project) in real-time during software development.
		
		In our experiments, comparing SourcererCC with the state-of-the-art 
		tools~\footnote{Deckard, CCFinder, NiCad, and iClones}, we found that it is the only clone detection tool to successfully 
		scale to 250 MLOC on a standard workstation with 12 GB RAM and efficiently detect the first three types of clones
		(precision 86\% and recall 86-100\%).
		Link to the demo: \url{https://youtu.be/l7F_9Qp-ks4}
		
	\end{abstract}
	
	
	
	
	\section{Introduction}
	
	Clone detection locates exact or similar pieces of code, known as clones, within or between software systems. Clones are created when developers reuse code by copy, paste and modify, although clones may be created by a number of other means~\cite{roy:queens:07}. Developers need to detect and manage their clones in order to maintain software quality, detect and prevent new bugs, reduce development risks, and costs~\cite{roy:queens:07}. Clone management and clone research studies depend on quality tools. 
	
	With the amount of source code increasing steadily, large-scale clone detection has become a necessity. Large-scale clone detection can be used for mining library candidates~\cite{ishihara}, detecting similar mobile applications~\cite{Chen:2014:android}, license violation detection~\cite{Koschke:CSMR:12}, reverse engineering product lines~\cite{hemel:2012wcre}, finding the provenance of a component~\cite{julius:2011:msr}, and code search~\cite{keivanloo:2011wcre,Kawaguchi:2009wcre}. Large-scale clone detection allows researchers to study cloning in large software ecosystems (e.g., Debian), or study cloning in open-source development communities like GitHub. Developers often clone modules or fork projects to meet the needs of different clients, and need the help of large-scale clone detectors to merge these cloned systems towards a product-line style of development. These applications require tools that scale to hundreds of millions of lines of code. However, very few tools/techniques can scale to the demands of clone detection in very large code bases~\cite{jeff_scalability2}. They achieve scalability by using deterministic~\cite{livieri:2007icse} or non-deterministic~\cite{jeff_scalability2} input partitioning and distributed execution of an existing non-scalable detector, using large distributed code indexes~\cite{hummel:icsm:2010}, or by comparing hashes after Type-1/2 normalization~\cite{ishihara}. These existing techniques have a number of limitations. The novel scalable algorithms~\cite{hummel:icsm:2010,ishihara} do not support Type-3 near-miss clones, where minor to significant editing activities might have taken place in the copy/pasted fragments, and therefore miss a large portion of the clones, since there are more Type-3 in the repositories than other types~\cite{Roy:2010:NFC:1779593.1779596}. Type-3 clones can be the most needed in large-scale clone detection applications~\cite{keivanloo:2011wcre,Chen:2014:android}. While input partitioning can scale existing non-scalable Type-3 detectors, this significantly increases the cumulative runtime, and requires distribution over a large cluster of machines to achieve scalability in absolute runtime~\cite{jeff_scalability2,livieri:2007icse}. Distributable tools~\cite{livieri:2007icse} are costly and difficult to setup and maintain. 
	
	To address this state of affairs, we developed a clone detection technique~\cite{sajnani-jsep,SourcererCC:2016}, which
	satisfies the following requirements: (1) accurate detection of near-miss clones, where minor to significant editing changes occur in the copy/pasted fragments; (2) programming language agnostic; (3) simple, non-distributed operation; and (4) scalability to hundreds of millions of lines of code using a standard workstation. This paper demonstrates the
	accompanying tool-suite (SourcererCC and SourcererCC-I) and several notable features that rely on the introduced approach in our earlier work. 
	
	\section{SourcererCC's Overview}
\begin{figure}
		\centering
		\includegraphics[scale=0.3]{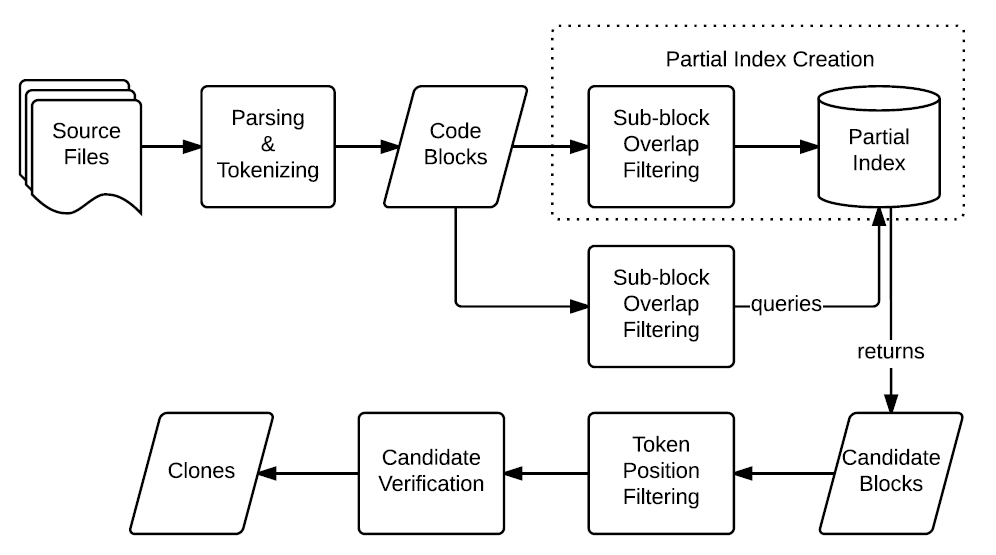} 
		\caption {SourcererCC's clone detection process}
		\label{fig:sourcererccoverview}
		\centering
	\end{figure}

	SourcererCC's general procedure is summarized in Figure~\ref{fig:sourcererccoverview}. It operates in two primary stages: (i) partial index creation; and (ii) clone detection.
	
	In the index creation phase, it parses the code blocks from the source files, and tokenizes them with a simple scanner that is aware of token and block semantics of a given language~\footnote{Currently, we have implemented this for Java, C and C\#, but can be easily extended to other languages}. From the code blocks it builds an inverted index mapping tokens to the blocks that contain them. Unlike previous approaches, it does not create an index of all tokens in the code blocks, instead it uses a filtering heuristic, named \textit{Sub-block Overlap Filtering} to construct a partial index of only a subset of the tokens in each block~\cite{SourcererCC:2016}.
	
	In the detection phase, SourcererCC iterates through all of the code blocks, retrieves their candidate clone blocks from the index. As per the filtering heuristic, only the tokens within the sub-block are used to query the index, which reduces the number of candidate blocks. After candidates are retrieved, SourcererCC uses another filtering heuristic named \textit{Token Position Filtering}, which exploits ordering of the tokens in a code block to measure a live upper-bound and lower-bound of similarity scores between the query and candidate blocks~\cite{SourcererCC:2016}. Candidates whose upper-bound falls below the similarity threshold are eliminated immediately without further processing. Similarly, candidates are accepted as soon as their lower-bound exceeds the similarity threshold. This is repeated until the clones of every code block are located.
	\section{Features}
	\label{section:feat}
	\subsection{Scalable}
	In our experiments, comparing SourcererCC with the state-
	of-the-art tools (Deckard, CCFinder, NiCad, and iClones), we found that it is the only clone detection tool to successfully scale to 250 MLOC on a standard
	workstation with 12 GB RAM~\footnote{SourcererCC successfully executed even on a machine with 8GB RAM} and detect first three types
	of clones~\cite{SourcererCC:2016}.
	\subsection{Accurate} 
	In our experiments, SourcererCC gave 100\% recall for Type 1 clones, 97-100\% recall for Type 2 clones and 86 - 99\% recall for Type 3 clones ~\cite{SourcererCC:2016}. It gave 86\% precision, a strong precision as per the literature~\cite{roy:queens:07, taxonomy}.
	\subsection{Language Independent}
	SourcererCC decouples the parsing logic from the indexing and clone detection logic, making it easier to support any number of languages as long as there is a parser for it. Currently, we have parsers to support Java, C, C++,C\#,and python.
	\subsection{Fast Index Creation and Clone Detection}
	SourcererCC uses filtering heuristics to index a small subset (for 70\% threshold, only 30\% of the block gets indexed) of a code block, making the size of the index small. Thus, indexing large software repositories takes considerably less time. Moreover, searching on smaller indexes is faster. For example, it took around 1.5 days to report clones on 100 MLOC, 3 times faster than CCFinderX, the second fastest tool~\cite{SourcererCC:2016}. 
	\subsection{Configurable}
	The threshold is configurable and users can increase or decrease it to find more stricter or liberal clones, respectively. Moreover, it is possible to detect clones at different granularities such as block level (statements between \{\}), method level, and file level. We note that we found the precision and recall to be optimum at 70\% threshold.
	\section{SourcererCC-I: Eclipse plug-in}
	\label{sec:plug}
	Many batch processing tools are designed for detecting clones in large software repositories. These tools, however, do not integrate well with the development process. 
	To this end, Lague et al. conducted a case study to assess the impact of integrating clone detection with the development process as a preventive control for maintenance issues~\cite{lague1997assessing}. They analyzed 89 million lines of code and found several opportunities where the integration of automated clone detection with the development process could have helped. 
	
	Although many tools have been developed to automatically detect clones, adoption of these tools during software development still faces three main challenges: 
	(i) Difficult to setup and configure; 
	(ii) Not designed to seamlessly integrate with the development process, as these are mostly stand alone software, designed for batch processing; and
	(iii) The output files of almost all clone detection tools is impossible to read by a human at any large-scale. 
	
	\subsection{SourcererCC-I Overview}
	SourcererCC-I addresses the above limitations. It is an Eclipse plug-in that instantaneously reports intra \& inter project method level clones in Java software. SourcererCC-I is built on top of SourcererCC, thus exploiting its powerful indexing and searching capabilities. SourcererCC-I detects the method a developer is currently working on, then pro-actively reports the clones of the method in a non-obtrusive manner. SourcererCC-I reports the clones inside the IDE, rather than just creating an output file that is a report. This bypasses the need to look at the text of a clone detection report file offering a great advantage to the user. 
	
	\subsection{SourcererCC-I's Architecture}
	
	\label{subsec:impl}
	
\begin{figure}[!htbp]
		\centering
		\includegraphics[scale=0.50]{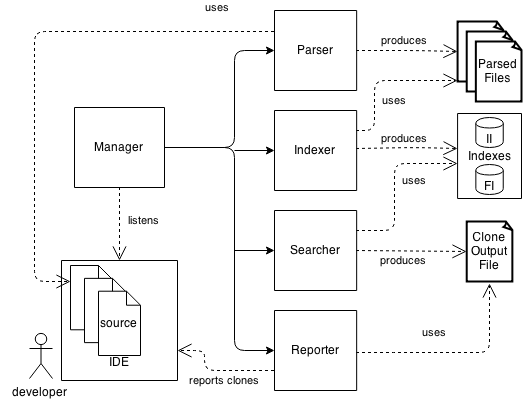}
		\caption{SourcererCC-I Architecture}
		\label{fig:architecture}
	\end{figure}

	SourcererCC-I has five modules as shown in Figure~\ref{fig:architecture}.
	
	\subsubsection{Manager: controls interaction between different modules} 
	The \textit{Manager} module acts as a controller, delegating jobs like \textit{create indexes}, \textit{update indexes},\textit{ search clones}, and \textit{report clones} to other modules. It mediates the data flow among them and listens to the \textit{change} and \textit{selection} events generated by the editor. On detection of such events, it performs action such as \textit{update indexes} or \textit{search clones}. 
	
	\subsubsection{Parser: parses projects and creates input for the indexer} 
	The \textit{Parser} generates input files for the \textit{Indexer} Module. On activation of the plug-in, the \textit{Manager} delegates the job of parsing the source files to the \textit{Parser} Module, which creates parsed files using Eclipse's JDT (Java Development Toolkit). 
	
	\subsubsection{Indexer: creates inverted and forward indexes of code blocks}
	The \textit{Indexer} uses the parsed files to create a partial inverted index and a forward index for each of the open projects in the Eclipse workspace. The partial inverted index is used to search the candidate clones whereas the forward index is used to verify if the candidates are clones or not. The more details on the creation and working of these indexes can be found in our other study~\cite{SourcererCC:2016}.
	
	\subsubsection{Searcher: searches clones in the indexes}
	The \textit{Manager} detects the current method and then creates its query block. The \textit{Searcher} uses this query block as an input and detect its clones using the indexes. It produces a clone output file containing meta information of clones.
	
	\subsubsection{Reporter: reports detected clones}
	After the \textit{Searcher} detects the clones of the current method, the \textit{Manager} asks the \textit{Reporter} module to perform two tasks i) create a marker on the editor (where the line numbers are written). This marker signifies the presence of clones of the current method in the project; and ii) display the list of clone methods using a view part in Eclipse. A user can quickly navigate to the clone methods by clicking any item in this list of clone methods.
	
	\subsection{SourcererCC-I's Features}
	\label{subsec:features}
	\vspace{.14cm}
	\noindent\textbf{1. Non-obtrusive user interface}
	
	SourcererCC-I has a minimalistic design in order to minimize the cognitive burden on the user.
	
	\noindent\textit{displaying clones.}
	The tool uses colored markers to notify the clones detected. A marker is either \textit{red}, \textit{blue}, or \textit{green} in color where
	\emph{red} signifies there are more than 10 clones of the current method in the project; \emph{yellow} signifies there are five to 10 clones; and
	\emph{green} signifies the existence of less than five clones
	
\begin{figure}[!htbp]
		\centering
		\includegraphics[scale=.45]{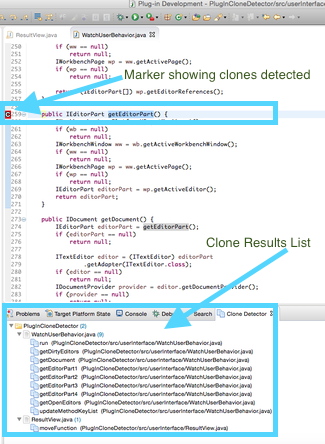}
		\caption\small{Screenshot of Eclipse, displaying that the plug-in has detected clones.}
		\label{fig:marker}
	\end{figure}

	Figure~\ref{fig:marker} shows a red colored marker (annotated inside the top blue rectangular region in the eclipse's editor) signifying that SourcererCC-I has found more than 10 clones of this method.
	
	\noindent\textit{navigating clones.}
	The output of most of the detection tools is impossible to read by a human at any large-scale. This is because of an extra step to link the output produce by the tools to the actual source of code fragments. To overcome this limitation, SourcererCC-I displays all the clones found for a method using their fully qualified name in the result view pane of Eclipse's console.  This not only facilitates easy navigation to cloned methods (in the editor) using a mouse click but also displays the clones in an hierarchical fashion using a tree structure. The detected clones are grouped by projects and also by files, making it easier for a developer to navigate. \\
	
	\noindent\textbf{2. Instant clone detection}
	
	SourcererCC's optimized index and filtering heuristics enable SourcererCC-I to detect clones in real time.\\
	
	\noindent\textbf{3. Incremental and fast index creation}
	
	SourcererCC-I index creation step is very fast i.e., few milliseconds for even project of 100,000 LOC. Thus, even for a project consisting of several thousands of lines of code, clone detection process can begin as soon as the project is loaded in the workspace.
	Moreover, the index creation is incremental, meaning, the index is not re-created for the whole project whenever a file or a method is changed. This is a very important feature for clone detection tools to be useful during ongoing development activities. \\
	
	\noindent\textbf{4. Inter-project and intra-project clone detection}
	
	SourcererCC-I has not only the ability to detect clones in a given project, but it can also easily detect clones across projects or repositories. \\
	
	\noindent\textbf{5. Quick and easy set-up}
	
	SourcererCC-I can be installed easily by following these 3 steps in the given order: (i) Open Eclipse and click \textit{Install New Software} in \textit{Help} menu; (ii) Click \textit{Add}. Then, in the \textit{Add Repository} window, enter the name SourcererCC-I, and in the location field enter the url \url{http://tinyurl.com/pm3xl2g}; and finally, (iii) follow the steps given in eclipse wizard to install the plug-in.

	\section{Related Tools}
	\label{sec:rel}
	We have extensively covered the work related to SourcererCC in our other paper~\cite{SourcererCC:2016}. In this section we focus on the tools which are related to SourcererCC-I.
	
	\noindent \textbf{Tools integrated with Development Environment.}
	Patricia Jablonski's proposed CnP, a tool to detect copy-and-paste clones in the IDE~\cite{hou2009cnp}. CnP establishes links between the original and the pasted clones and uses their content information later on for error detection and other purposes. Unlike our tool, CnP doesn't use a clone detector to detect clones and hence is not capable of finding legacy clones. Moreover, CnP cannot find accidental clones.
	
	CP-Miner detects copy-paste errors in the context of traditional clone detection~\cite{li2004cp}. It uses a token-based approach with data mining and a gap constraint. We are using an index-based token comparison technique, which falls under the information retrieval approach. Also, unlike our tool, CP-Miner is not available freely and one needs to buy a license to use it. 
	
	CloneTracker, an Eclipse plug-in is a tool that keeps track of the evolution of clones~\cite{duala2008clonetracker}. For it's input, it relies on the output of a clone detector tool, that needs to be run before hand to generate all the clones in a system. CloneTracker, unlike SourcererCC-I, does not detects the clones in real time.

	\noindent\textbf{Stand Alone Tools.}
	There are some stand alone tools, namely Dup~\cite{baker1995finding}, iClones~\cite{gode2009incremental}, CCFinder~\cite{kamiya:2002zr}, and Nicad~\cite{roy2008nicad} that do batch processing to detect clones, but these tools are not designed to be integrated with the development process. Nicad, however, can be integrated to an IDE but being a batch processing tool it lacks the capability to detect clones on the fly.
	CloneDR, a commercial AST based clone detection tool~\cite{baxter1998clone} detects clones by finding identical subtrees. CloneDR claims to have Eclipse integration for IBM Enterprise languages.
	
	%
	%

	\section{Conclusion and Future Work}
	\label{sec:conc}
	
	In this paper we presented: (i) SourcererCC, a token-based clone detector
	that targets the first three clone types, and exploits an index
	to achieve scalability to large inter-project repositories
	using a standard workstation. SourcererCC uses an optimized
	inverted-index to quickly query the potential clones
	of a given code block. Filtering heuristics based on token ordering
	are used to significantly reduce the size of the index,
	the number of code-block comparisons needed to detect the
	clones, as well as the number of required token-comparisons
	needed to judge a potential clone.
	(ii) SourcererCC-I, an easy-to-use, free, and well packaged clone detection plug-in for the Eclipse IDE that can detect clones on the fly. The tool pro-actively finds method level clones (inter and intra-project) and reports them in a non-obtrusive manner. 
	In future we plan to exploit the architecture of both Sourcerer-CC and SourcererCC-I, which decouples the parsing logic from the clone detection logic, to add support for more languages and different granularities of code blocks.
	
	\noindent\textbf{Tool Artifacts.}
	Link to install the Eclipse plug-in: \url{http://mondego.ics.uci.edu/projects/clonedetection/tool/latest} \& link to the source of SourcererCC:
	\url{http://mondego.ics.uci.edu/projects/clonedetection/tool/source}
	\bibliographystyle{ieeetr}
	\bibliography{bib}  
\end{document}